%% file: main.tex
\documentclass[journal=accacs,manuscript=article,layout=twocolumn]{achemso}
\usepackage[version=3]{mhchem} 

\usepackage{multirow}
\usepackage{mdframed}
\usepackage{rotating}
\usepackage{booktabs}
\usepackage{etoolbox}
\usepackage{enumitem}
\usepackage{subcaption}
\usepackage[nonumberlist]{glossaries}
\mciteErrorOnUnknownfalse
\usepackage{xcolor}
\usepackage{comment}
\usepackage{array}
\usepackage{hyperref}
\usepackage[version=3]{mhchem}
\newcounter{magicrownumbers}
\preto\tabular{\setcounter{magicrownumbers}{0}}
\usepackage{todonotes}
\usepackage{longtable}

\newacronym{DFT}{DFT}{Density Functional Theory}
\newacronym{QE}{QE}{Quantum Espresso}
\newacronym{VASP}{VASP}{Vienna Ab initio Simulation Package}
\newacronym{S2EF}{\textit{S2EF}}{Structure to Energy and Forces}
\newacronym{IS2RE}{\textit{IS2RE}}{Initial Structure to Relaxed Energy}
\newacronym{IS2RS}{\textit{IS2RS}}{Initial Structure to Relaxed Structure}
\newacronym{OC20}{OC20}{Open Catalyst 2020}
\newacronym{ML}{ML}{Machine Learning}
\newacronym{TM}{TM}{Transition Metal}
\newacronym{FCC}{FCC}{Face-centered cubic}
\newacronym{MKM}{MKM}{Microkinetic Model}
\newacronym{MAE}{MAE}{Mean Absolute Error}

\glsaddall

\newcommand{\cmu}{Department of Chemical Engineering, Carnegie Mellon University}
\newcommand{\scott}{Scott Institute for Energy Innovation, Carnegie Mellon University}

\title[]
  {Catlas: an automated framework for catalyst discovery demonstrated for direct syngas conversion}

\author{Brook Wander}
\affiliation{\cmu}
\author{Kirby Broderick}
\affiliation{\cmu}
\author{Zachary W. Ulissi}

\affiliation{\scott}
\alsoaffiliation{\cmu}
\email{zulissi@andrew.cmu.edu}
\abbreviations{}

\let\oldmaketitle\maketitle
\let\maketitle\relax

\begin{document}



\twocolumn[
\begin{@twocolumnfalse}
\oldmaketitle

\begin{abstract}

Catalyst discovery is paramount to support access to energy and key chemical feedstocks in a post fossil fuel era. Exhaustive computational searches of large material design spaces using ab-initio methods like density functional theory (DFT) are infeasible. We seek to explore large design spaces at relatively low computational cost by leveraging large, generalized, graph-based machine learning (ML) models, which are pretrained and therefore require no upfront data collection or training. We present catlas, a framework that distributes and automates the generation of adsorbate-surface configurations and ML inference of DFT energies to achieve this goal. Catlas is open source, making ML assisted catalyst screenings easy and available to all. To demonstrate its efficacy, we use catlas to explore catalyst candidates for the direct conversion of syngas to multi-carbon oxygenates. For this case study, we explore 947 stable/ metastable binary, transition metal intermetallics as possible catalyst candidates. On this subset of materials, we are able to predict the adsorption energy of key descriptors, *CO and *OH, with near-DFT accuracy (0.16, 0.14 eV MAE, respectively). Using the projected selectivity towards C2+ oxygenates from an existing microkinetic model, we identified 144 candidate materials. For 10 promising candidates, DFT calculations reveal a good correlation with our assessment using ML. Among the top elemental combinations were Pt-Ti, Pd-V, Ni-Nb, and Ti-Zn, all of which appear unexplored experimentally.  

\end{abstract}
\end{@twocolumnfalse}
]

\input{sections/introduction}

\input{sections/methods}

\input{sections/results_discussion}

\input{sections/conclusion}

\begin{acknowledgement}
Acknowledgment is made to the Donors of the American Chemical Society Petroleum Research Fund for support (or partial support) of this research. Kirby Broderick acknowledges research support by the Department of Energy, Office of Science, Office of Basic Sciences, Data Science for Knowledge Discovery for Chemical and Materials Research program, under award DE-SC0020392. We would like to acknowledge Janice Lan from Meta AI Research for running and processing the OC20-like *CO data that was used here to make Figure \ref{fig:parity}a and assess model performance on *CO.
\end{acknowledgement}
\clearpage
\bibliography{bib}
\clearpage

\end{document}

%% file: sections/introduction.tex
\section{Introduction}
As we look to recast our energy infrastructure, we need to discover novel catalysts to support our new vision. Catalysts will allow access to key chemical feedstocks we have come to rely on today, but from renewable resources. One method of accessing carbon feedstocks is conversion of synthesis gas (syngas). Synthesis gas is a mixture of carbon monoxide and hydrogen. It may be accessed from renewable biomass feedstocks\cite{prospects2017} or through electrochemical reduction of carbon dioxide and water\cite{co2rr}. The conversion of syngas may be done directly, which could prove to be an attractive approach due to its relative simplicity. Still, controlling the product distribution and selectivity is an important outstanding challenge. There is particular interest in selectively accessing multi-carbon oxygenates because of their utility and high market value\cite{co2rr_review}.

Development of catalysts for the direct conversion of syngas to multi-carbon oxygenates has primarily been focused on 4 classes of materials: Rh-based, Mo-based, modified Fischer-Tropsch, and modified methanol synthesis catalysts\cite{prospects2017, Recent_Adv_CO_direct}. Despite being the subject of investigation for nearly a century, no economically viable catalysts have been identified to pursue commercial implementation\cite{prospects2017}. The only elemental catalyst that has shown some selectivity towards multi-carbon oxygenates is Rh\cite{prospects2017, ethanol_biomass2007, review_ethanol2008}. Experimental and theoretical studies had shown that under-coordinated surfaces have low selectivity towards C2+ oxygenates\cite{review_ethanol2008} and that less active terrace sites have some selectivity towards the desired products\cite{schumann_selectivity}. A more recent work has shown that the previous treatment of adsorbate-adsorbate interactions on Rh (111) surfaces was inadequate and arrived at the opposite conclusion for Rh\cite{new_ads_ads}, so the understanding of this chemistry is still evolving. To form multi-carbon oxygenates, the catalyst must be simultaneously good at disassociating *CO to form *\ce{CH_x} intermediates on the surface and able to maintain *CO such that it may be inserted into the *\ce{CH_x} intermediates\cite{cao, rh_details, Recent_Adv_CO_direct}. Striking a balance between these two competing reactions makes finding a catalyst for this chemistry difficult. Copper cobalt binary intermetallics have been studied. Cobalt based catalysts are industrially used for Fischer-Tropsch (FT) synthesis\cite{prospects2017}. They efficiently form *\ce{CH_x} intermediates and facilitate C-C coupling to form hydrocarbon products. It has been suggested that CuCo materials work synergistically to make multi-carbon oxygenates\cite{cao}. Cobalt sites disassociate *CO, while copper sites maintain *CO, thereby allowing it to be inserted into the *\ce{CH_x} intermediates. We are interested in binary intermetallics because they have not been thoroughly explored for this chemistry.

\begin{figure}[ht]
    \centering
    \includegraphics[width=0.5\textwidth]{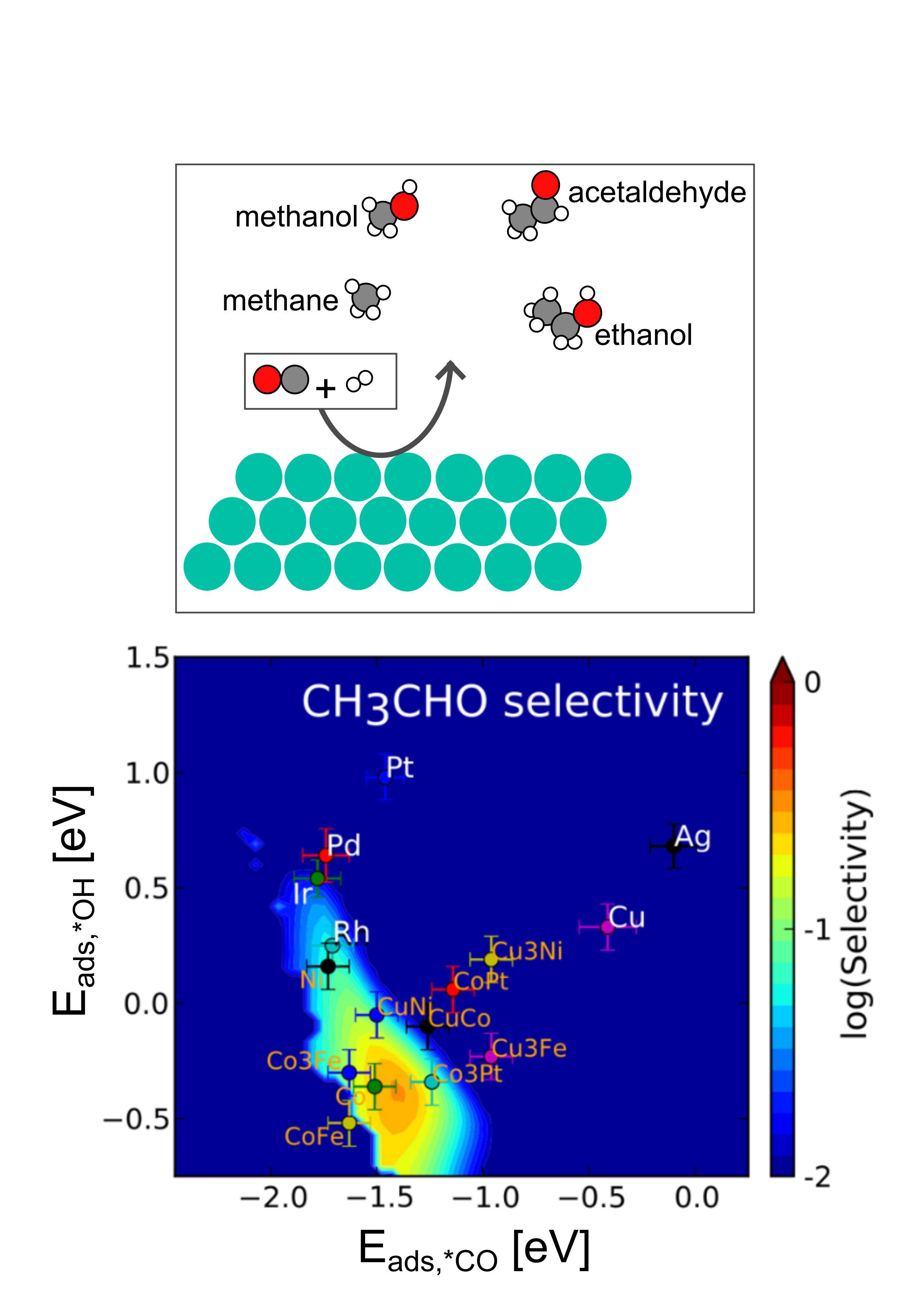}
    \vspace{-30pt}
    \caption{The acetaldehyde selectivity as a function of *CO and *OH adsorption energies as solved by Schumann et al.\cite{schumann_selectivity} which also considered methanol, methane, and ethanol as alternate products. (bottom) Reproduced with permission from ref. 7 Copyright 2018, ACS}
    \label{fig:mkm_summary}
\end{figure}

To run a screening, a clear objective is necessary (i.e. to achieve high selectivity and/or activity). The computational approach to catalyst discovery necessitates the observation of many different sites, on many different surfaces, for every material in order to understand how it might behave as a catalyst and therefore assess it against the objective. One binary material would require hundreds of DFT calculations to fully uncover its behavior as a catalyst. Observation of adsorbed intermediate and transition state energies allow an understanding of the overall rate and overall selectivity to be developed through microkinetic modeling. Microkinetic models (MKMs) are well established for determining trends\cite{mkm_fundementals}. This is done in practice\cite{SA-MoS2, kinetics_H_X, CO2rr_c-c_couple}, but is very costly to scale. When considering many materials, it is tractable to use descriptor-based objectives where just a few adsorption energies are assigned target values\cite{HTS_w_ML_2017, Tran2018, 2d_mat_HER, DA-NF}. Because dense site exploration of large design spaces with DFT is not feasible, ML has been of interest to accelerate this process. Most efforts have required the generation of domain specific datasets and bespoke ML models \cite{Review_ML_universal_models, HTS_w_ML_2017, SA_W_ML} to accomplish this. Without ML, screenings focus on specific classes of materials such a single atom catalysts\cite{SA-MoS2}, dual-atom catalysts\cite{DA-NF}, or 2D materials\cite{2d_mat_HER} where the design space can be more well-bounded and brute force is more feasible. Both of these approaches are computationally intensive. 

Here, we used off-the-shelf pretrained graph neural network models, which required no upfront DFT data collection or model training, to explore 947 binary intermetallics as catalyst candidates. Looking at facets with miller indices not exceeding 1, we considered 16k surfaces, which yield 1 million adsorbate-surface configurations. To facilitate this process we developed and introduce a software package, catlas, which automates the process of generating slabs from bulk structures, placing adsorbates on the slabs, and performing inference on the adsorbate-surface configurations. Catlas orchestrates the distribution of these trivially parallelizable tasks so that many GPUs and/or CPUs may be used to accelerate the process.  We then used an existing microkinetic model to project the selectivity as a function of two descriptor adsorption energies. The intermediate and transition state energies were correlated with the energies of *CO and *OH on the (111) surfaces of Face-centered cubic (FCC) pure metals to project the selectivity across a complex reaction network. The resulting selectivity heatmap (Figure \ref{fig:mkm_summary}), reveals that materials which adsorb *OH with adsorption energies in the range of 0: -0.75 eV and adsorb *CO with energies in the range of -1.25: -1.75 eV may show selectivity towards multi-carbon oxygenates. We use this conclusion for the (111) surface of pure metals and extrapolate to (111)-like surfaces of binary intermetallics. ML models were used to predict the adsorption energies of *OH and *CO for 947 binary intermetallics from the Materials Project\cite{MP} with an energy above hull not exceeding 0.1 eV/atom. The predicted adsorption energies were used to classify whether the materials could be interesting as potential catalyst candidates. We found 144 promising materials which have been ranked in our analysis.

%% file: sections/methods.tex
\begin{figure*}
    \centering
    \includegraphics[width=\textwidth]{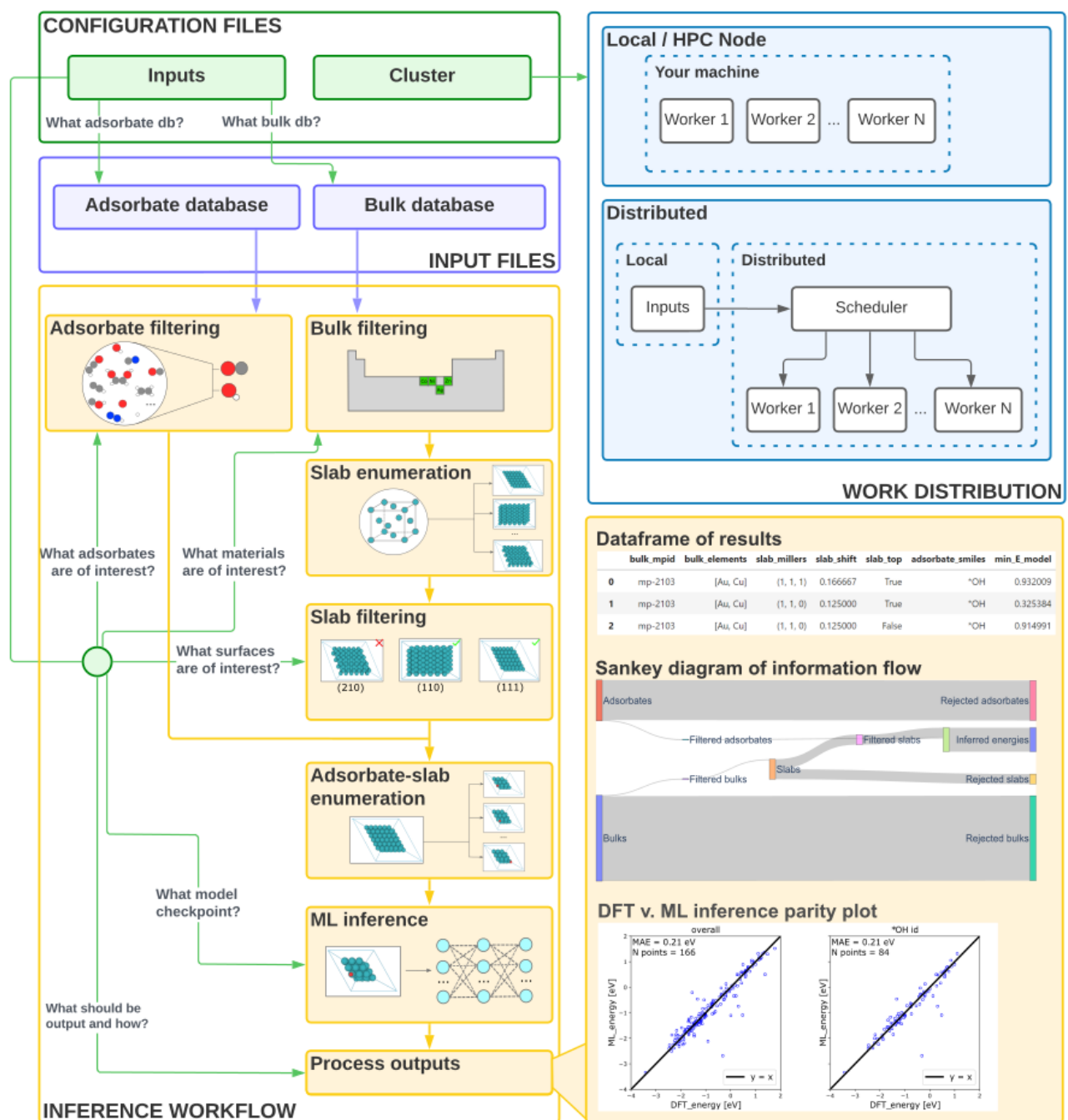}
        \vspace{-30pt}
    \caption{The architecture of catlas: an open source framework for adsorbate-surface configuration generation and ML inference. Details show the options for distributing work, the workflow for enumerating adsorbate-surface configurations and performing inference, run outputs, and how user specified configurations interact with the architecture.}
    \vspace{-30pt}
    \label{fig:catlas_workflow}
\end{figure*}

\section{Methods}

\subsection{Model Selection}

All models considered were graph neural networks trained on the OC20 dataset\cite{OC20}. Graph neural networks are well-suited for learning chemical properties because they capture connectivity of atoms within their architecture without the need for specially engineered features\cite{CGCNN}. Atoms are represented as nodes in the graphs and edges represent interactions between atoms. The OC20 dataset contains unary, binary, and ternary materials made up of 55 different elements. The dataset contains many transition metal systems, but it also contains non-metals, metalloids, post-transition, alkali, and alkali earth metals. It covers 82 different adsorbates comprised of C, H, O, and N, with 1-12 atoms. The OC20 dataset was not catered to materials discovery for the direct conversion of syngas. *CO does not even appear in the training set, so the models have not even seen an example of *CO on a surface. There is opportunity to apply this method to broader classes of materials and other chemistries using the same models. Of the graph neural network architectures that have been trained on the OC20 dataset, we considered GemNet\cite{gemnet} models because they are currently state-of-the-art amongst models with pretrained weights publicly available according to the Open Catalyst Project leaderboard. GemNet leverages directed edge embeddings and edge-based message passing to achieve predictions which are invariant to translation and equivariant to rotation as atomic forces and energies should be.

\begin{figure*}
    \centering
    \includegraphics[width=0.95\textwidth]{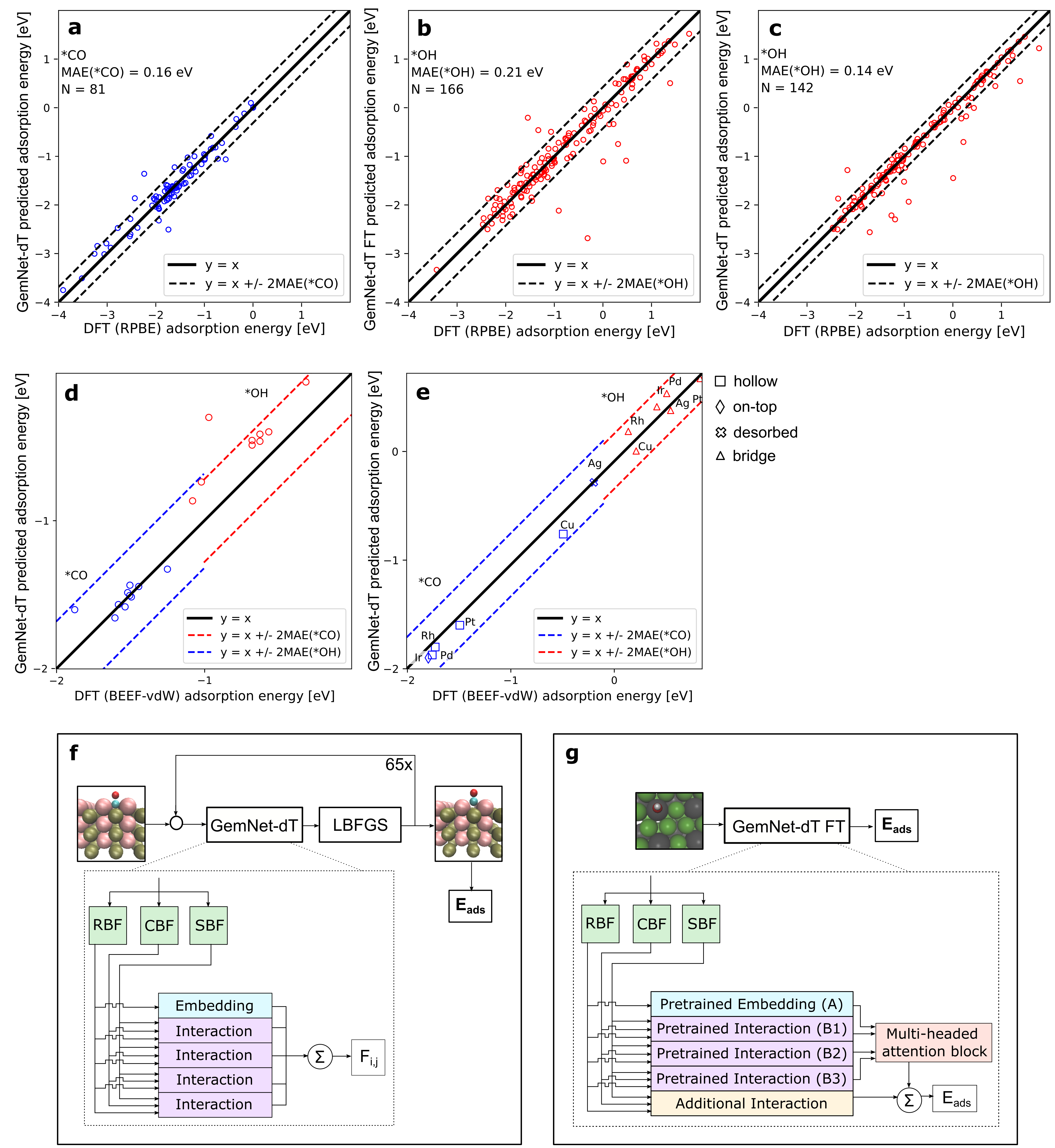}
    \caption{Parity plots used to assess model performance on the domain of interest. (a-b) *CO and *OH on binary, transition metal intermetallics, respectively. DFT calculations were performed with identical functional, psuedopotentials, and settings as the model training data. (c) *CO and *OH on the (111) surface of unary transition metals. The DFT calculations, taken from Schumann et al., were used to build scaling relations for the MKM considered\cite{schumann_selectivity}. (d-e) Model details for *CO and *OH inference, respectively.}
    \label{fig:parity}
\end{figure*}

Model selection comes with a trade-off: lower mean absolute error (MAE) between model predicted values and DFT calculated values often requires higher computational cost. Inference where the adsorption energy is directly predicted from the initial structure (initial structure to relaxed energy - IS2RE - direct) is substantially cheaper than using ML to iteratively optimize the structure and predict the energy from the relaxed structure (structure to energy and forces - S2EF). These two schemes are shown in Figure \ref{fig:parity}. *OH has been included in the training data, and binary, transition metal (TM) intermetallics make up a large portion of the training data. For these reasons, direct approaches give reasonable performance for *\ce{OH} for this use case. The best performing model is a GemNet-dT model that was initially trained to perform iterative relaxation steps, and was finetuned to directly predict the relaxed energy\cite{AdeeshTL} (GemNet-dT FT). This data was used as an initial filtering step. For any surface where the ML predicted *OH adsorption energy fell in the range [-1.2,-0.5] eV, the *OH ML inference was repeated using lower error relaxation model (GemNet-dT)\cite{gemnet}. This domain comfortably bounds the domain of interest from the selectivity heatmap. Inference was also performed on this subset of surfaces for *CO using the exact same GemNet-dT\cite{gemnet} pretrained model. For *CO, 65 relaxation steps was selected because it minimized the MAE with respect to DFT validation data. For the relaxed *OH inference, 136 steps was similarly selected.

\subsection{Candidate Classification}
Catlas was used to enumerate adsorbate-surface configurations for all surfaces with miller indices not exceeding one. Any relaxations with adsorbate desorption, adsorbate dissociation, or surface reconstruction were discarded. For each of the enumerated surfaces, once dense inference had been performed, the minimum binding energies for each descriptor (*CO and *OH) were computed and taken to be the representative energy for that surface. Analysis was performed on a per surface basis. No surface energy calculations were performed so it is not guaranteed that surfaces with good descriptor energies will appear experimentally or that their properties will dominate observed behavior. Because the MKM was developed for face-centered cubic (FCC) (111) surfaces, surfaces were discarded if they were not (111)-like (close packed). The selectivity information was determined by resolving the microkinetic model presented by Schumann et al. There were small differences in the selectivity map solved for this work, but the overall domain of interest was the same. The potential candidacy of binary transition metal combinations was considered using two criteria: (1) the minimum distance of any surface to the maximum observed selectivity towards acetaldehyde. The distance from the maximum selectivity was calculated for all surfaces. For each elemental combination, the minimum distance across all surfaces was calculated and was taken to represent the composition. (2) We counted the number of surfaces which lie in a descriptor space containing a projected selectivity greater than or equal to 0.1. If any part of the descriptor space bounded by an ellipse with radii equal to twice the mean absolute error (MAE) of the ML models has a selectivity greater than 0.1, then that surface was considered to be a candidate. This is repeated for all surfaces and the number of candidates for each metal combination was calculated. MAE was used as a proxy for uncertainty because uncertainty information is not available from the models.

\subsection{Catlas}
The architecture of catlas is summarized in Figure \ref{fig:catlas_workflow}. The workflow starts by filtering which bulk materials and adsorbates should be considered. Adsorbates are simply filtered by their SMILES strings. Bulks may be filtered by many criteria: Pourbaix stability, elemental composition, number of unique elements, material identifier, energy above hull, band gap, and bulk size. Certain elements may be specified as required to, for example, look at nitrides. Additionally, the paradigm of active-host materials may be selected for. From the materials of interest, slabs are enumerated using pymatgen\cite{pymatgen}. The slabs may be filtered by their size and a maximum miller index if desired. Next, adsorbate-surface configurations are enumerated using CatKit\cite{catkit}. Both slab and adsorbate-slab enumerations rely on the data generation infrastructure that was created for the OC20 dataset\cite{OC20} and therefore create OC20-like adsorbate-surface configurations which may aptly be used with ML models trained on the dataset. Finally, ML inference is performed on each of the adsorbate-surface configurations. 

Three important outputs are generated in a catlas run: a summary of results (pandas dataframe), Sankey diagram, and parity plots. The summary dataframe memorializes all of the key information from the run: the bulk, the adsorbate, the information that uniquely defines each surface (miller indices, shift, top of the slab or bottom), which model was used, the inferred energies, etcetera. The Sankey diagram summarizes the enumeration and inference process by showing what survives filtering steps and how many objects are created at enumeration steps. The parity plots inform model performance on a per adsorbate basis for the types of material considered. 

Catlas uses dask\cite{dask} to distribute its inference workflow. Dask-distributed has backends for parallelization on local machines, standard HPC clusters, and kubernetes clusters, as well as support for GPU workloads. In this work, we successfully scaled CPU workloads across 20 nodes with 600 cores and GPU workloads to 2 Nodes with 8 GPUs, but this is not a limit. A run of calculations commences simply by specifying the design space to be explored in an input yaml file and executing the repository's main file. By default, the adsorbates and materials that may be considered are those contained within the OC20 dataset\cite{OC20}, but this is configurable if additional bulk database and adsorbate database files are added. The OC20 dataset sampled a subset of materials from the Materials Project\cite{MP} which contain 3 or fewer unique elements (unary, binary, ternary) and only contain a specific set of elements. The time required to complete calculations is dependent on the materials. For our case study, using GemNet-dT to relax adsorbate-surface systems, approximately 50 surfaces (1,200 adsorbate-surface configurations) may be considered per hour using 1 GPU. Using GemNet-dT FT to directly predict the adsorption energies, approximately 10 surfaces (200 adsorbate-surface configurations) may be considered per hour using one 4-core worker. This means that at scale, it took just 10 hours to complete direct calculations on 16k surfaces (500k calculations) and 16 hours to calculate 6k surfaces (150k calculations).
\begin{figure*}[h]
    \centering
    \includegraphics[width=\textwidth]{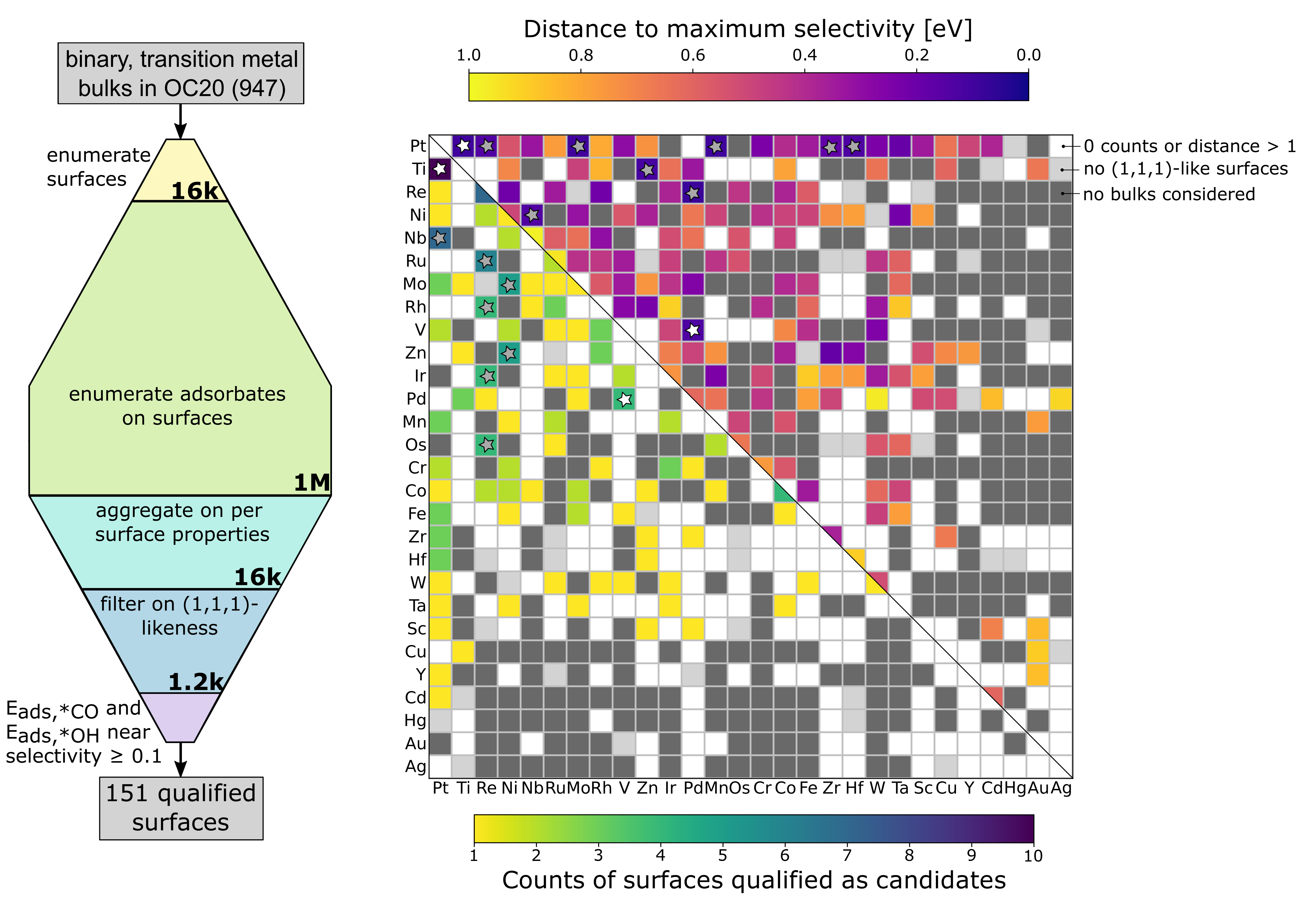}
    \caption{(left) The expansion and contraction of information through enumeration and filtering in this work. (right) Grid summarizing element combinations of interest or the direct conversion of syngas to C${_2+}$ oxygenates. The lower wedge shows the number of surfaces per combination that were classified as hits by proximity to a selectivity greater than 0.1 (approach 2) and the upper wedge shows the minimum distance to the maximum selectivity for each combination (approach 1).}
    \label{fig:analysis}
\end{figure*}

%% file: sections/results_discussion.tex
\section{Results and Discussion}
  \begin{figure*}[ht!]
    \centering
    \includegraphics[width=0.95\textwidth]{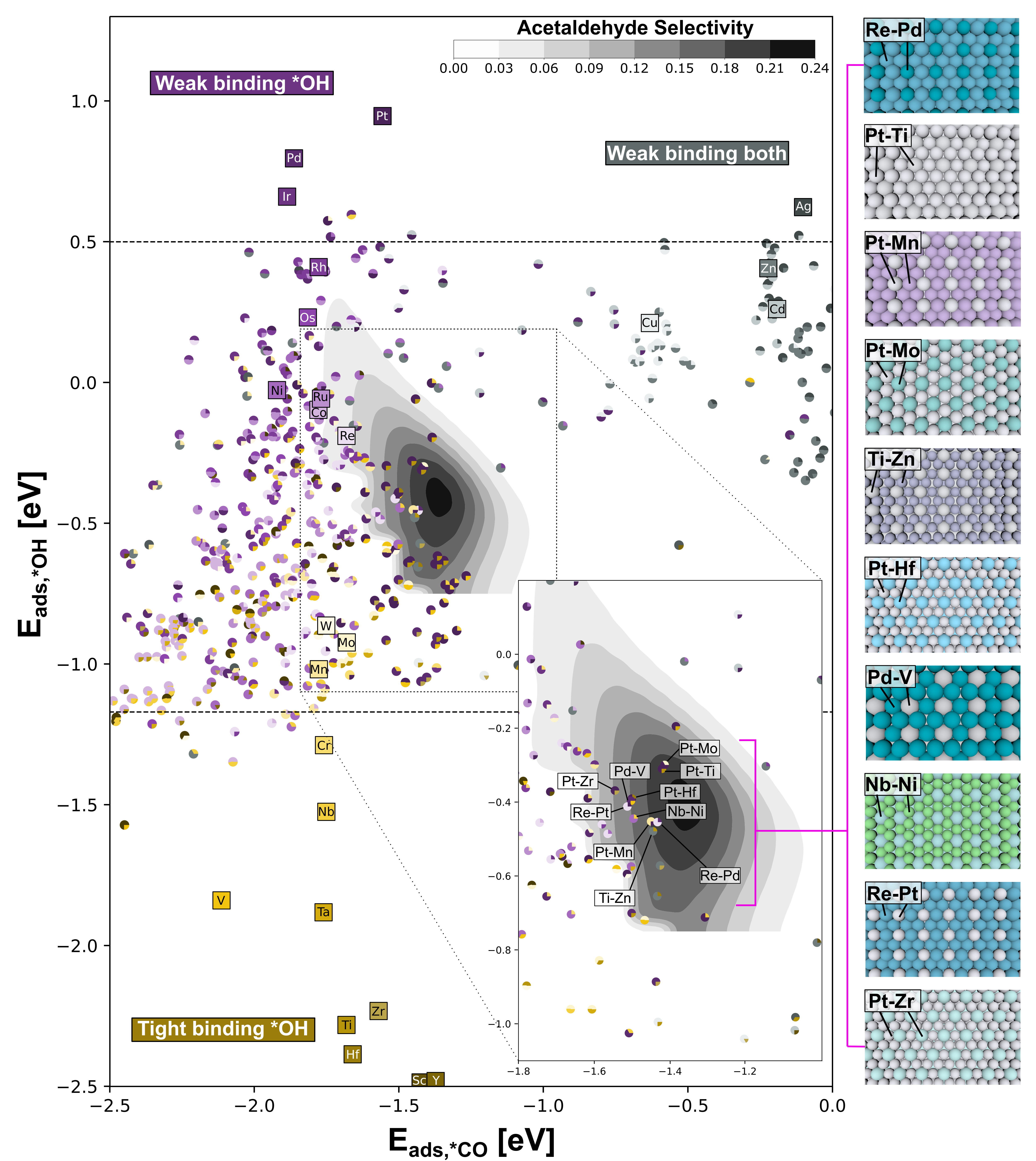}
    \caption{The minimum binding *CO and *OH binding energies per (111)-like surface overlaid onto the MKM acetaldehyde selectivity heatmap developed by Schumann et al.\cite{schumann_selectivity} and resolved for this work. Surfaces are shown as pie charts which have proportions representing the bulk composition. Elements have been colored according to whether they lie in the weak binding regime for both descriptors (grey), weak binding for *OH only (purple), or tight binding for both (yellow). Inset shows detailed location of the top surfaces per composition near the peak selectivity. The top 10 surfaces are pictured to the right.}
    \label{fig:descriptor_space_view}
\end{figure*}

 Parity plots comparing the GemNet ML predicted energies and DFT energies are shown in Figure \ref{fig:parity}. The domain specific MAE for *OH using the direct GemNet-dT FT model was 0.21 eV. The domain specific MAE for *CO and *OH using the GemNet-dT model were 0.16 eV and 0.14 eV, respectively. There is good coverage of binary intermetallic benchmark data over the range of adsorption energies considered for the MKM (-2.45: 0.25 eV for *CO and -0.75: 1.50 eV for *OH). The annotation of twice the MAE reveals its utility in our candidate classification approach. The majority (roughly 90\%) of data lies within these bounds. Figure \ref{fig:parity}a-c only considers the difference between the GemNet model and corresponding (RPBE) DFT calculations. 
 
 There are still two important sources of error to consider: (1) differences between the actual intermediate and transition state energies and those propagated via scaling relations implicitly in the MKM and (2) differences between the DFT approach employed in the data used to train the GemNet models and the approach used to collect data for the MKM. Differences between the actual intermediate and transition state energies and those propagated via scaling relations implicitly in the MKM are relevant here because we are extrapolating the MKM to binary intermetallics with similar facets despite it being developed for unary surfaces. Although this approach will not perfectly treat all possibilities, it should treat the majority of cases well, and therefore is a useful tool for broad exploration.

 Figure \ref{fig:parity}d-e provides some insight into differences caused by the difference in functionals employed. OC20 data was calculated using the RPBE functional, while the  MKM data used BEEF-vdW. These two functionals have been shown to have similar performance for short range chemisorption energy interactions\cite{dft_bench} like those considered in this work. Overall, there is good agreement between the OC20 trained, GemNet-dT predicted energies and the MKM DFT data (Fig \ref{fig:parity}e). The GemNet-dT predicted energies in Figure \ref{fig:parity}d-e are the minimum energies per surface, where many heuristically placed sites were considered per surface. For the MKM DFT data, specific sites were considered. All calculations were performed on bridge sites for *OH and most calculations were performed at hollow sites for *CO. Figure \ref{fig:parity}d compares the minimum energy per surface predicted by GemNet-dT and the minimum energy per surface calculated by BEEF-vdW DFT for this work. There is good agreement for *CO. There is also a good correlation for *OH, but there is an offset of about 0.2 eV, which was not observed in Fig. \ref{fig:parity}e. We suspect this offset was not observed in Figure \ref{fig:parity}d because of a cancellation of errors. Likely, this is primarily caused by the *OH BEEF-vdW DFT values being systematically lower than the RPBE DFT values, but the MKM considers higher energy bridge sites. Because there is good agreement between the unary MKM data and the GemNet-dT predicted energies, we did not perform any corrections to our inferred energies. Still, we considered the implications of systematic biases in our data by applying an offset and observing differences in the outcome of our analysis (Table \ref{tab:sensitivity}). If we had applied a correction of -0.2 eV to the *OH values, 81\% of our classifications would be retained and we would have missed just 16 candidate surfaces (88\% would have been found by our approach). In general, our analysis is relatively insensitive to biases in *OH energies, and relatively sensitive to biases in *CO energies.

\begin{table*}[ht]
    \begin{tabular}{p{4cm}p{1.6cm}p{1.6cm}p{1.6cm}p{1.6cm}p{1.6cm}p{1.6cm}}
        \hline
        \multirow{2}{*}{Applied error}&
            \multicolumn{2}{c}{Correctly classified}&
            \multicolumn{2}{c}{False negative}&
            \multicolumn{2}{c}{False positive}\\
        &N &\% & N &\% & N & \% \\
        \hline
        *OH + 0.1 eV & 151 & 100 & 0  & 0 & 0  & 0\\
        *OH + 0.2 eV & 151 & 100 & 0  & 0 & 0 & 0\\
        *OH - 0.1 eV & 133 & 88  & 11 & 8 & 18 & 12\\
        *OH - 0.2 eV & 123 & 81  & 16 & 12& 28 & 19\\
        *CO + 0.1 eV & 150 & 99  & 55 & 27 & 1  & 1\\
        *CO + 0.2 eV & 146 & 97  & 92 & 39 & 5  & 3\\
        *CO - 0.1 eV & 103 & 68  & 6  & 6 & 48 & 32\\
        *CO - 0.2 eV &  71 & 47  & 10 & 12 & 80 & 53\\

        \hline
    \end{tabular}
    \caption{Analysis sensitivity to biases in GemNet-dT predictions.}
\label{tab:sensitivity}
\end{table*}

\begin{table*}[ht]
    \begin{tabular}{p{4cm}p{1.6cm}p{1.6cm}p{1.6cm}p{1.6cm}p{1.6cm}p{1.6cm}}
        \hline
        \multirow{2}{*}{Nuclearities}&
            \multicolumn{2}{c}{All surfaces}&
            \multicolumn{2}{c}{(111)-like}&
            \multicolumn{2}{c}{(111)-like, candidate}\\
        &N &\% & N &\% & N & \% \\
        \hline
        monomer, other & 6,911 & 43.2 & 680 & 59.3 & 119 & 78.8\\
        dimer, other & 1,383 & 8.7 & 7& 0.6 & 0 & 0.0\\
        trimer, other & 748 & 4.7 & 7 & 0.6 & 0 & 0.5\\
        4+, other & 1,007 & 6.3 & 1 & 0.1 & 0 & 0.5\\
        infinite, 0 & 1,061 & 6.6 & 213 & 18.6 & 13 & 8.6\\
        both semi-finite & 3,348 & 20.9 & 203 & 17.7 & 17 & 11.3\\
        non-finite (other) & 1,622 & 10.1 & 41 & 3.6 & 2 & 1.3 \\
        \hline
        all & 15,981 & 100.0 & 1,147 & 100.0 & 151 & 100.0\\

        \hline
    \end{tabular}
    \caption{Summary of surface motifs for all enumerated slabs considered.}
\label{tab:motifs}
\end{table*}

The results of assessing candidates by the minimum distance to the maximum selectivity and the number of surfaces near high selectivity are summarized in Figure \ref{fig:analysis}. Enumeration yields 1 million unique structures to be considered to elucidate possible catalysts among binary TMs. By filtering on (111)-likeness, just 7\% of the surfaces remain. Just 1\% of the original pool is near the domain of interest and (111)-like. The 10 closest element combinations to the maximum selectivity projected by the MKM have been noted with stars in Figure \ref{fig:analysis}. Similarly, elemental combinations with 4 or more surfaces that were classified as hits are noted with stars. White stars have been used where combinations satisfied both of these criteria, while grey stars have been used where one was satisfied. The ten combinations with minimum descriptor energies closest to the maximum (Re-Pd, Pt-Ti, Pt-Mn, Pt-Mo, Ti-Zn, Pt-Hf, Pd-V, Nb-Ni, Re-Pt, Pt-Zr) are previously unexplored as potential catalyst candidates reported in literature.

It is difficult to make comparisons between the results here and the best known materials because materials reported are highly modified\cite{prospects2017, syndir_review, Recent_Adv_CO_direct}. For example, top Rh-based catalysts are promoted with Mn, Li, and Fe. The intricacy of engineered supports and modifications are beyond the scope of what may be captured here. Mo-based catalysts are covalent materials (Mo-S, Mo-C, Mo-O, or Mo-P), which were outside of the scope of this screening as well. Modified methanol synthesis and Fischer-Tropsch catalyst both use known catalysts for those chemistries as a starting place, but are highly modified from there\cite{prospects2017}. Even the simple modified-FT example (Cu-Co or Cu-Fe) may not be compared because there are no CuCo or CuFe materials in the Materials Project database\cite{MP} that meet the stability criteria of less than 0.1 eV/atom above the convex hull.
\begin{figure}[h]
    \centering
    \includegraphics[width=0.45\textwidth]{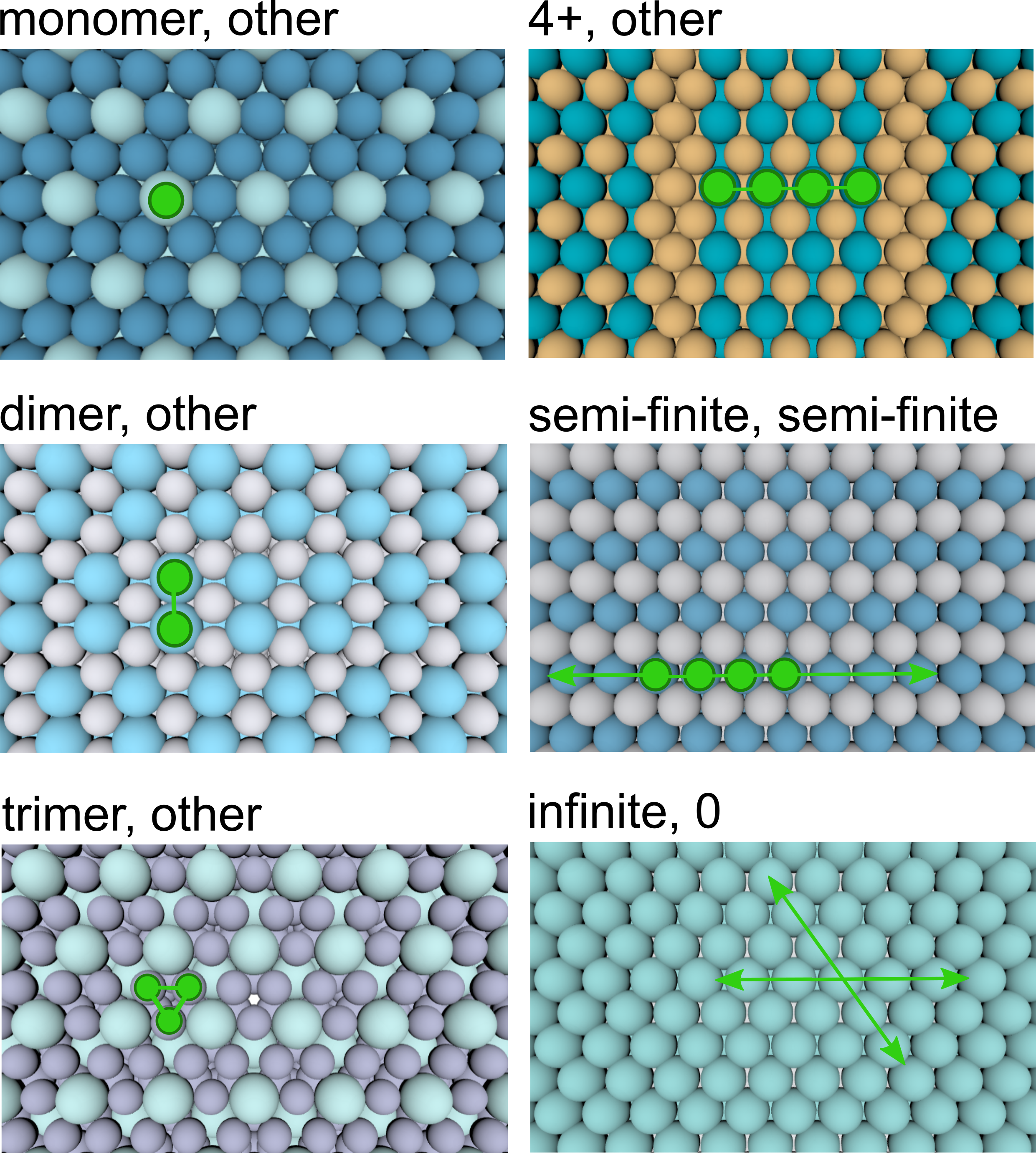}
    \caption{Categories of surface motifs identified, shown here on surfaces that have also been categorized as (111)-like.}
    \label{fig:motifs}
\end{figure}
A descriptor-space view may be seen in Figure \ref{fig:descriptor_space_view}. The paradigm of tight and weak binding is expressed by coloring elements by their binding characteristics (grey - weak binding for both descriptors, purple - tight binding for *CO and weak binding for *OH, yellow - tight binding both).  The dotted lines show the bounds of *OH values which were discarded in the initial course screening step using the less accurate GemNet-dT FT model. Most of the best surfaces are an elemental mixture of one pure metal that weakly binds *OH and one that tightly binds *OH, as would be expected by a simple tie-line analysis. Of the ten surfaces closest to the maximum, all except those that are Re containing are a tight-weak binding mixture. The minimum energies of 111-like surfaces for the materials considered sparsely populate the domain of interest where acetaldehyde selectivity is projected to be high. Pure iron is not shown here because its descriptor energies are very tight binding (*OH: -2.4 eV , *CO: -3.6 eV), but alloy surfaces containing it are appropriately colored with very dark yellow.
      
The proportions of slab surface motifs for the 16,000 surfaces considered here are shown in Table \ref{tab:motifs}. Examples of the nuclearity types making up these motifs may be seen in Figure \ref{fig:motifs}. Those with nuclearities of three or more may have different surface connectivities (i.e. linear or triangular for three). The nuclearity types were calculated using the graph based algorithm developed by Sharma et al.\cite{nuclearity} Surfaces where one element type has no neighbors of the same element (monomers) are 43.2\% of those enumerated. Once filtered by 111-likeness, monomers make up a majority of the data (59.3\%), and filtering by candidate classification makes this majority stronger (78.8\%). All of the ten surfaces closest to the maximum show this nuclearity. Despite making up 20\% of the data, there were no dimer+ candidates. Infinite nuclearity in one element and zero in another indicates that the top layer only contains a single unique element. Semi-finite is infinite in one direction. These two nuclearities make up essentially the rest of the candidate pool. Non-finite (other) captures all other surfaces with non-integer nuclearities for both elements.

%% file: sections/conclusion.tex
\section{Conclusion}
Exploring large design spaces for novel catalysts using DFT is infeasible. DFT surrogates have historically required the construction of expensive domain-specific DFT training data sets. Here, we show that off-the-shelf models pretrained on the OC20\cite{OC20} dataset have achieved near DFT accuracy for a broad domain of materials captured in the dataset. We present catlas, an open source framework to parallelize and automate the process of evaluating materials as catalysts based upon descriptors. We demonstrate its efficacy by considering the direct conversion of syngas to multi-carbon oxygenates. With the well-posed objective for *OH and *CO descriptor energies from Schumann et al., we were able to screen 947 stable and meta-stable materials and discover a subset which may be well positioned for this chemistry. With the design space narrowed by our approach, candidates may be prioritized and tractably assessed using more detailed studies with DFT or by experimental investigation. The utility of catlas is underscored by the evolving understanding of the kinetics for the direct conversion of syngas to multi-carbon oxygenates. When an improved descriptor-based kinetic model is available, catlas may be used to assess potential catalyst candidates with ease. The generalized framework of catlas, makes it extensible. With clear objective descriptors, catlas may be used to discover catalysts for any arbitrary chemistry.

%% file: main.bbl
\providecommand{\latin}[1]{#1}
\makeatletter
\providecommand{\doi}
  {\begingroup\let\do\@makeother\dospecials
  \catcode`\{=1 \catcode`\}=2 \doi@aux}
\providecommand{\doi@aux}[1]{\endgroup\texttt{#1}}
\makeatother
\providecommand*\mcitethebibliography{\thebibliography}
\csname @ifundefined\endcsname{endmcitethebibliography}
  {\let\endmcitethebibliography\endthebibliography}{}
\begin{mcitethebibliography}{32}
\providecommand*\natexlab[1]{#1}
\providecommand*\mciteSetBstSublistMode[1]{}
\providecommand*\mciteSetBstMaxWidthForm[2]{}
\providecommand*\mciteBstWouldAddEndPuncttrue
  {\def\EndOfBibitem{\unskip.}}
\providecommand*\mciteBstWouldAddEndPunctfalse
  {\let\EndOfBibitem\relax}
\providecommand*\mciteSetBstMidEndSepPunct[3]{}
\providecommand*\mciteSetBstSublistLabelBeginEnd[3]{}
\providecommand*\EndOfBibitem{}
\mciteSetBstSublistMode{f}
\mciteSetBstMaxWidthForm{subitem}{(\alph{mcitesubitemcount})}
\mciteSetBstSublistLabelBeginEnd
  {\mcitemaxwidthsubitemform\space}
  {\relax}
  {\relax}

\bibitem[Luk \latin{et~al.}(2017)Luk, Mondelli, Ferré, Stewart, and
  Pérez-Ramírez]{prospects2017}
Luk,~H.~T.; Mondelli,~C.; Ferré,~D.~C.; Stewart,~J.~A.; Pérez-Ramírez,~J.
  Status and prospects in higher alcohols synthesis from syngas. \emph{Chem.
  Soc. Rev.} \textbf{2017}, \emph{46}, 1358--1426\relax
\mciteBstWouldAddEndPuncttrue
\mciteSetBstMidEndSepPunct{\mcitedefaultmidpunct}
{\mcitedefaultendpunct}{\mcitedefaultseppunct}\relax
\EndOfBibitem
\bibitem[Perathoner and Centi(2014)Perathoner, and Centi]{co2rr}
Perathoner,~S.; Centi,~G. CO2 recycling: a key strategy to introduce green
  energy in the chemical production chain. \emph{ChemSusChem} \textbf{2014},
  \emph{7}, 1274--1282\relax
\mciteBstWouldAddEndPuncttrue
\mciteSetBstMidEndSepPunct{\mcitedefaultmidpunct}
{\mcitedefaultendpunct}{\mcitedefaultseppunct}\relax
\EndOfBibitem
\bibitem[Nitopi \latin{et~al.}(2019)Nitopi, Bertheussen, Scott, Liu, Engstfeld,
  Horch, Seger, Stephens, Chan, Hahn, Nørskov, Jaramillo, and
  Chorkendorff]{co2rr_review}
Nitopi,~S.; Bertheussen,~E.; Scott,~S.~B.; Liu,~X.; Engstfeld,~A.~K.;
  Horch,~S.; Seger,~B.; Stephens,~I. E.~L.; Chan,~K.; Hahn,~C.;
  Nørskov,~J.~K.; Jaramillo,~T.~F.; Chorkendorff,~I. Progress and Perspectives
  of Electrochemical CO2 Reduction on Copper in Aqueous Electrolyte.
  \emph{Chemical Reviews} \textbf{2019}, \emph{119}, 7610--7672, PMID:
  31117420\relax
\mciteBstWouldAddEndPuncttrue
\mciteSetBstMidEndSepPunct{\mcitedefaultmidpunct}
{\mcitedefaultendpunct}{\mcitedefaultseppunct}\relax
\EndOfBibitem
\bibitem[Damma and Smirniotis(2021)Damma, and Smirniotis]{Recent_Adv_CO_direct}
Damma,~D.; Smirniotis,~P.~G. Recent advances in the direct conversion of syngas
  to oxygenates. \emph{Catal. Sci. Technol.} \textbf{2021}, \emph{11},
  5412--5431\relax
\mciteBstWouldAddEndPuncttrue
\mciteSetBstMidEndSepPunct{\mcitedefaultmidpunct}
{\mcitedefaultendpunct}{\mcitedefaultseppunct}\relax
\EndOfBibitem
\bibitem[Spivey and Egbebi(2007)Spivey, and Egbebi]{ethanol_biomass2007}
Spivey,~J.~J.; Egbebi,~A. Heterogeneous catalytic synthesis of ethanol from
  biomass-derived syngas. \emph{Chem. Soc. Rev.} \textbf{2007}, \emph{36},
  1514--1528\relax
\mciteBstWouldAddEndPuncttrue
\mciteSetBstMidEndSepPunct{\mcitedefaultmidpunct}
{\mcitedefaultendpunct}{\mcitedefaultseppunct}\relax
\EndOfBibitem
\bibitem[Subramani and Gangwal(2008)Subramani, and Gangwal]{review_ethanol2008}
Subramani,~V.; Gangwal,~S.~K. A Review of Recent Literature to Search for an
  Efficient Catalytic Process for the Conversion of Syngas to Ethanol.
  \emph{Energy \& Fuels} \textbf{2008}, \emph{22}, 814--839\relax
\mciteBstWouldAddEndPuncttrue
\mciteSetBstMidEndSepPunct{\mcitedefaultmidpunct}
{\mcitedefaultendpunct}{\mcitedefaultseppunct}\relax
\EndOfBibitem
\bibitem[Schumann \latin{et~al.}(2018)Schumann, Medford, Yoo, Zhao, Bothra,
  Cao, Studt, Abild-Pedersen, and Nørskov]{schumann_selectivity}
Schumann,~J.; Medford,~A.~J.; Yoo,~J.~S.; Zhao,~Z.-J.; Bothra,~P.; Cao,~A.;
  Studt,~F.; Abild-Pedersen,~F.; Nørskov,~J.~K. Selectivity of Synthesis Gas
  Conversion to C2+ Oxygenates on fcc(111) Transition-Metal Surfaces. \emph{ACS
  Catalysis} \textbf{2018}, \emph{8}, 3447--3453\relax
\mciteBstWouldAddEndPuncttrue
\mciteSetBstMidEndSepPunct{\mcitedefaultmidpunct}
{\mcitedefaultendpunct}{\mcitedefaultseppunct}\relax
\EndOfBibitem
\bibitem[Deimel \latin{et~al.}(0)Deimel, Prats, Seibt, Reuter, and
  Andersen]{new_ads_ads}
Deimel,~M.; Prats,~H.; Seibt,~M.; Reuter,~K.; Andersen,~M. Selectivity Trends
  and Role of Adsorbate–Adsorbate Interactions in CO Hydrogenation on Rhodium
  Catalysts. \emph{ACS Catalysis} \textbf{0}, \emph{0}, 7907--7917\relax
\mciteBstWouldAddEndPuncttrue
\mciteSetBstMidEndSepPunct{\mcitedefaultmidpunct}
{\mcitedefaultendpunct}{\mcitedefaultseppunct}\relax
\EndOfBibitem
\bibitem[Cao \latin{et~al.}(2018)Cao, Schumann, Wang, Zhang, Xiao, Bothra, Liu,
  Abild-Pedersen, and Nørskov]{cao}
Cao,~A.; Schumann,~J.; Wang,~T.; Zhang,~L.; Xiao,~J.; Bothra,~P.; Liu,~Y.;
  Abild-Pedersen,~F.; Nørskov,~J.~K. Mechanistic Insights into the Synthesis
  of Higher Alcohols from Syngas on CuCo Alloys. \emph{ACS Catalysis}
  \textbf{2018}, \emph{8}, 10148--10155\relax
\mciteBstWouldAddEndPuncttrue
\mciteSetBstMidEndSepPunct{\mcitedefaultmidpunct}
{\mcitedefaultendpunct}{\mcitedefaultseppunct}\relax
\EndOfBibitem
\bibitem[Choi and Liu(2009)Choi, and Liu]{rh_details}
Choi,~Y.; Liu,~P. Mechanism of Ethanol Synthesis from Syngas on Rh(111).
  \emph{Journal of the American Chemical Society} \textbf{2009}, \emph{131},
  13054--13061, PMID: 19702298\relax
\mciteBstWouldAddEndPuncttrue
\mciteSetBstMidEndSepPunct{\mcitedefaultmidpunct}
{\mcitedefaultendpunct}{\mcitedefaultseppunct}\relax
\EndOfBibitem
\bibitem[Motagamwala and Dumesic(2021)Motagamwala, and
  Dumesic]{mkm_fundementals}
Motagamwala,~A.~H.; Dumesic,~J.~A. Microkinetic Modeling: A Tool for Rational
  Catalyst Design. \emph{Chemical Reviews} \textbf{2021}, \emph{121},
  1049--1076, PMID: 33205961\relax
\mciteBstWouldAddEndPuncttrue
\mciteSetBstMidEndSepPunct{\mcitedefaultmidpunct}
{\mcitedefaultendpunct}{\mcitedefaultseppunct}\relax
\EndOfBibitem
\bibitem[Yang \latin{et~al.}(2020)Yang, Song, Zhou, Wang, Chi, Shen, Yang, and
  Feng]{SA-MoS2}
Yang,~T.; Song,~T.~T.; Zhou,~J.; Wang,~S.; Chi,~D.; Shen,~L.; Yang,~M.;
  Feng,~Y.~P. High-throughput screening of transition metal single atom
  catalysts anchored on molybdenum disulfide for nitrogen fixation. \emph{Nano
  Energy} \textbf{2020}, \emph{68}, 104304\relax
\mciteBstWouldAddEndPuncttrue
\mciteSetBstMidEndSepPunct{\mcitedefaultmidpunct}
{\mcitedefaultendpunct}{\mcitedefaultseppunct}\relax
\EndOfBibitem
\bibitem[MacQueen \latin{et~al.}(2021)MacQueen, Royko, Crandall, Heyden,
  Pagán-Torres, and Lauterbach]{kinetics_H_X}
MacQueen,~B.; Royko,~M.; Crandall,~B.~S.; Heyden,~A.; Pagán-Torres,~Y.~J.;
  Lauterbach,~J. Kinetics Study of the Hydrodeoxygenation of Xylitol over a
  ReOx-Pd/CeO2 Catalyst. \emph{Catalysts} \textbf{2021}, \emph{11}\relax
\mciteBstWouldAddEndPuncttrue
\mciteSetBstMidEndSepPunct{\mcitedefaultmidpunct}
{\mcitedefaultendpunct}{\mcitedefaultseppunct}\relax
\EndOfBibitem
\bibitem[Kuo \latin{et~al.}(2021)Kuo, Chou, Shen, Hong, Chao, Lu, and
  Cheng]{CO2rr_c-c_couple}
Kuo,~T.-C.; Chou,~J.-W.; Shen,~M.-H.; Hong,~Z.-S.; Chao,~T.-H.; Lu,~Q.;
  Cheng,~M.-J. First-Principles Study of C–C Coupling Pathways for CO2
  Electrochemical Reduction Catalyzed by Cu(110). \emph{The Journal of Physical
  Chemistry C} \textbf{2021}, \emph{125}, 2464--2476\relax
\mciteBstWouldAddEndPuncttrue
\mciteSetBstMidEndSepPunct{\mcitedefaultmidpunct}
{\mcitedefaultendpunct}{\mcitedefaultseppunct}\relax
\EndOfBibitem
\bibitem[Li \latin{et~al.}(2017)Li, Wang, Chin, Achenie, and
  Xin]{HTS_w_ML_2017}
Li,~Z.; Wang,~S.; Chin,~W.~S.; Achenie,~L.~E.; Xin,~H. High-throughput
  screening of bimetallic catalysts enabled by machine learning. \emph{J.
  Mater. Chem. A} \textbf{2017}, \emph{5}, 24131--24138\relax
\mciteBstWouldAddEndPuncttrue
\mciteSetBstMidEndSepPunct{\mcitedefaultmidpunct}
{\mcitedefaultendpunct}{\mcitedefaultseppunct}\relax
\EndOfBibitem
\bibitem[Tran and Ulissi(2018)Tran, and Ulissi]{Tran2018}
Tran,~K.; Ulissi,~Z.~W. Active learning across intermetallics to guide
  discovery of electrocatalysts for CO2 reduction and H2 evolution.
  \emph{Nature Catalysis} \textbf{2018}, \emph{1}, 696--703\relax
\mciteBstWouldAddEndPuncttrue
\mciteSetBstMidEndSepPunct{\mcitedefaultmidpunct}
{\mcitedefaultendpunct}{\mcitedefaultseppunct}\relax
\EndOfBibitem
\bibitem[Yang \latin{et~al.}(2020)Yang, Zhou, Song, Shen, Feng, and
  Yang]{2d_mat_HER}
Yang,~T.; Zhou,~J.; Song,~T.~T.; Shen,~L.; Feng,~Y.~P.; Yang,~M.
  High-Throughput Identification of Exfoliable Two-Dimensional Materials with
  Active Basal Planes for Hydrogen Evolution. \emph{ACS Energy Letters}
  \textbf{2020}, \emph{5}, 2313--2321\relax
\mciteBstWouldAddEndPuncttrue
\mciteSetBstMidEndSepPunct{\mcitedefaultmidpunct}
{\mcitedefaultendpunct}{\mcitedefaultseppunct}\relax
\EndOfBibitem
\bibitem[Lv \latin{et~al.}(2021)Lv, Wei, Huang, Dai, and Frauenheim]{DA-NF}
Lv,~X.; Wei,~W.; Huang,~B.; Dai,~Y.; Frauenheim,~T. High-Throughput Screening
  of Synergistic Transition Metal Dual-Atom Catalysts for Efficient Nitrogen
  Fixation. \emph{Nano Letters} \textbf{2021}, \emph{21}, 1871--1878, PMID:
  33587621\relax
\mciteBstWouldAddEndPuncttrue
\mciteSetBstMidEndSepPunct{\mcitedefaultmidpunct}
{\mcitedefaultendpunct}{\mcitedefaultseppunct}\relax
\EndOfBibitem
\bibitem[Sulley and Montemore(2022)Sulley, and
  Montemore]{Review_ML_universal_models}
Sulley,~G.~A.; Montemore,~M.~M. Recent progress towards a universal machine
  learning model for reaction energetics in heterogeneous catalysis.
  \emph{Current Opinion in Chemical Engineering} \textbf{2022}, \emph{36},
  100821\relax
\mciteBstWouldAddEndPuncttrue
\mciteSetBstMidEndSepPunct{\mcitedefaultmidpunct}
{\mcitedefaultendpunct}{\mcitedefaultseppunct}\relax
\EndOfBibitem
\bibitem[Zafari \latin{et~al.}(2020)Zafari, Kumar, Umer, and Kim]{SA_W_ML}
Zafari,~M.; Kumar,~D.; Umer,~M.; Kim,~K.~S. Machine learning-based high
  throughput screening for nitrogen fixation on boron-doped single atom
  catalysts. \emph{J. Mater. Chem. A} \textbf{2020}, \emph{8}, 5209--5216\relax
\mciteBstWouldAddEndPuncttrue
\mciteSetBstMidEndSepPunct{\mcitedefaultmidpunct}
{\mcitedefaultendpunct}{\mcitedefaultseppunct}\relax
\EndOfBibitem
\bibitem[Jain \latin{et~al.}(2013)Jain, Ong, Hautier, Chen, Richards, Dacek,
  Cholia, Gunter, Skinner, Ceder, and Persson]{MP}
Jain,~A.; Ong,~S.~P.; Hautier,~G.; Chen,~W.; Richards,~W.~D.; Dacek,~S.;
  Cholia,~S.; Gunter,~D.; Skinner,~D.; Ceder,~G.; Persson,~K.~a. {The Materials
  Project: A materials genome approach to accelerating materials innovation}.
  \emph{APL Materials} \textbf{2013}, \emph{1}, 011002\relax
\mciteBstWouldAddEndPuncttrue
\mciteSetBstMidEndSepPunct{\mcitedefaultmidpunct}
{\mcitedefaultendpunct}{\mcitedefaultseppunct}\relax
\EndOfBibitem
\bibitem[Chanussot \latin{et~al.}(2021)Chanussot, Das, Goyal, Lavril, Shuaibi,
  Riviere, Tran, Heras-Domingo, Ho, Hu, Palizhati, Sriram, Wood, Yoon, Parikh,
  Zitnick, and Ulissi]{OC20}
Chanussot,~L. \latin{et~al.}  Open Catalyst 2020 (OC20) Dataset and Community
  Challenges. \emph{ACS Catalysis} \textbf{2021}, \emph{11}, 6059--6072\relax
\mciteBstWouldAddEndPuncttrue
\mciteSetBstMidEndSepPunct{\mcitedefaultmidpunct}
{\mcitedefaultendpunct}{\mcitedefaultseppunct}\relax
\EndOfBibitem
\bibitem[Xie and Grossman(2018)Xie, and Grossman]{CGCNN}
Xie,~T.; Grossman,~J.~C. Crystal Graph Convolutional Neural Networks for an
  Accurate and Interpretable Prediction of Material Properties. \emph{Phys.
  Rev. Lett.} \textbf{2018}, \emph{120}, 145301\relax
\mciteBstWouldAddEndPuncttrue
\mciteSetBstMidEndSepPunct{\mcitedefaultmidpunct}
{\mcitedefaultendpunct}{\mcitedefaultseppunct}\relax
\EndOfBibitem
\bibitem[Gasteiger \latin{et~al.}(2021)Gasteiger, Becker, and
  G\"{u}nnemann]{gemnet}
Gasteiger,~J.; Becker,~F.; G\"{u}nnemann,~S. GemNet: Universal Directional
  Graph Neural Networks for Molecules. Advances in Neural Information
  Processing Systems. 2021; pp 6790--6802\relax
\mciteBstWouldAddEndPuncttrue
\mciteSetBstMidEndSepPunct{\mcitedefaultmidpunct}
{\mcitedefaultendpunct}{\mcitedefaultseppunct}\relax
\EndOfBibitem
\bibitem[Kolluru \latin{et~al.}(2022)Kolluru, Shoghi, Shuaibi, Goyal, Das,
  Zitnick, and Ulissi]{AdeeshTL}
Kolluru,~A.; Shoghi,~N.; Shuaibi,~M.; Goyal,~S.; Das,~A.; Zitnick,~C.~L.;
  Ulissi,~Z. Transfer learning using attentions across atomic systems with
  graph neural networks (TAAG). \emph{The Journal of Chemical Physics}
  \textbf{2022}, \emph{156}, 184702\relax
\mciteBstWouldAddEndPuncttrue
\mciteSetBstMidEndSepPunct{\mcitedefaultmidpunct}
{\mcitedefaultendpunct}{\mcitedefaultseppunct}\relax
\EndOfBibitem
\bibitem[Ong \latin{et~al.}(2013)Ong, Richards, Jain, Hautier, Kocher, Cholia,
  Gunter, Chevrier, Persson, and Ceder]{pymatgen}
Ong,~S.~P.; Richards,~W.~D.; Jain,~A.; Hautier,~G.; Kocher,~M.; Cholia,~S.;
  Gunter,~D.; Chevrier,~V.~L.; Persson,~K.~A.; Ceder,~G. {Python Materials
  Genomics (pymatgen): A robust, open-source python library for materials
  analysis}. \emph{Computational Materials Science} \textbf{2013}, \emph{68},
  314--319\relax
\mciteBstWouldAddEndPuncttrue
\mciteSetBstMidEndSepPunct{\mcitedefaultmidpunct}
{\mcitedefaultendpunct}{\mcitedefaultseppunct}\relax
\EndOfBibitem
\bibitem[Boes \latin{et~al.}(2019)Boes, Mamun, Winther, and Bligaard]{catkit}
Boes,~J.~R.; Mamun,~O.; Winther,~K.; Bligaard,~T. Graph Theory Approach to
  High-Throughput Surface Adsorption Structure Generation. \emph{The Journal of
  Physical Chemistry A} \textbf{2019}, \emph{123}, 2281--2285, PMID:
  30802053\relax
\mciteBstWouldAddEndPuncttrue
\mciteSetBstMidEndSepPunct{\mcitedefaultmidpunct}
{\mcitedefaultendpunct}{\mcitedefaultseppunct}\relax
\EndOfBibitem
\bibitem[{Dask Development Team}(2016)]{dask}
{Dask Development Team}, Dask: Library for dynamic task scheduling. 2016\relax
\mciteBstWouldAddEndPuncttrue
\mciteSetBstMidEndSepPunct{\mcitedefaultmidpunct}
{\mcitedefaultendpunct}{\mcitedefaultseppunct}\relax
\EndOfBibitem
\bibitem[Mallikarjun~Sharada \latin{et~al.}(2019)Mallikarjun~Sharada, Karlsson,
  Maimaiti, Voss, and Bligaard]{dft_bench}
Mallikarjun~Sharada,~S.; Karlsson,~R. K.~B.; Maimaiti,~Y.; Voss,~J.;
  Bligaard,~T. Adsorption on transition metal surfaces: Transferability and
  accuracy of DFT using the ADS41 dataset. \emph{Phys. Rev. B} \textbf{2019},
  \emph{100}, 035439\relax
\mciteBstWouldAddEndPuncttrue
\mciteSetBstMidEndSepPunct{\mcitedefaultmidpunct}
{\mcitedefaultendpunct}{\mcitedefaultseppunct}\relax
\EndOfBibitem
\bibitem[Subramani and Gangwal(2008)Subramani, and Gangwal]{syndir_review}
Subramani,~V.; Gangwal,~S.~K. A Review of Recent Literature to Search for an
  Efficient Catalytic Process for the Conversion of Syngas to Ethanol.
  \emph{Energy \& Fuels} \textbf{2008}, \emph{22}, 814--839\relax
\mciteBstWouldAddEndPuncttrue
\mciteSetBstMidEndSepPunct{\mcitedefaultmidpunct}
{\mcitedefaultendpunct}{\mcitedefaultseppunct}\relax
\EndOfBibitem
\bibitem[Sharma \latin{et~al.}(2022)Sharma, Nguyen, Janik, and
  Ulissi]{nuclearity}
Sharma,~U.; Nguyen,~A.; Janik,~M.; Ulissi,~Z. Site Geometry as a Descriptor for
  Catalyst Selectivity in Intermetallics. pre-print, 2022; Article under
  review. Citation will be updated when available\relax
\mciteBstWouldAddEndPuncttrue
\mciteSetBstMidEndSepPunct{\mcitedefaultmidpunct}
{\mcitedefaultendpunct}{\mcitedefaultseppunct}\relax
\EndOfBibitem
\end{mcitethebibliography}
